\documentclass[useAMS,usenatbib]{mn2e}

\usepackage{graphicx}

\newcommand{\dxdy}[2]{{\frac{\partial{#1}}{\partial{#2}}}}
\newcommand{\dxdys}[2]{{\frac{\partial^{2}{#1}}{\partial{#2}^{2}}}}
\newcommand{\DxDy}

\title[Gravity Wave Driven Flows]{Nonlinear Dynamics of Gravity Wave Driven Flows in the Solar Radiative Interior}
\author[T.M. Rogers, K.M. MacGregor and G.A. Glatzmaier]{T.M. Rogers$^{1}$\thanks{E-mail:
trogers@ucar.edu}, K.B. MacGregor$^{2}$ and G.A. Glatzmaier$^{3}$\footnotemark[1]\\
$^{1}$Astronomy and Astrophysics Postdoctoral Fellow, High Altitude Observatory, NCAR Boulder, CO 80301\\
$^{2}$High Altitude Observatory, NCAR Boulder, CO 80301\\
$^{3}$Department of Earth and Planetary Sciences, University of California, Santa Cruz, CA, 95064}

\begin{document}
\maketitle

\begin{abstract}
We present results of nonlinear numerical simulations of gravity wave driven shear flow oscillations in the equatorial plane of the solar radiative interior.  These results show that many of the assumptions of quasi-linear theory are not valid.  When only two waves are forced (prograde and retrograde) oscillatory mean flow is maintained; but critical layers often form and are dynamically important.  When a spectrum of waves is forced, the non-linear wave-wave interactions are dynamically important, often acting to ${\it decrease}$ the maintenance of a mean flow.  The (in)coherence of such wave-wave interactions must be taken into account when describing wave driven mean flows.  
\end{abstract}

\section{Introduction}

Explaining the solar internal rotation remains an outstanding problem in stellar physics.  Seismic observations indicate that the entire convection zone of the Sun rotates differentially, with the equator spinning faster than the poles, similar to what is observed at the solar surface \citep{th96,th03}.  This differential rotation persists to the base of the solar convection zone, below which the angular velocity transitions to approximately uniform rotation.  The transition between the differential rotation of the solar convection zone and the uniform rotation of the radiative interior occurs in an unresolved layer known as the solar tachocline \citep{sz92}.  

The differential rotation of the convection zone has been modeled in three dimensional simulations of the solar convection zone \citep{mbt06}, including the effect of an imposed subadiabatic tachocline as proposed by \cite{re05}.  However, in order to understand the maintenance of the tachocline, one needs to understand the maintenance of the uniform rotation of the solar radiative interior.

Currently, there are two main theories for the uniform rotation of the radiative interior: one magnetohydrodynamic and one purely hydrodynamic in nature.   The former argues that a primordial poloidal magnetic field confined to the radiative interior could enforce uniform rotation via the Lorentz force as described by Ferraro's isorotation law \citep{fer37}.  This model is able to maintain uniform rotation only if the field lines are completely confined to the radiative interior \citep{mc99}.  If field lines cross the convective-radiative interface the differential rotation of the convection zone is communicated into the radiative interior, contrary to helioseismic inferences.   It has been suggested that meridional circulation driven in the convection zone might penetrate the tachocline and keep the field confined \citep{gm98}.  Whereas some models have success with this approach \citep{kr06}, other models fail \citep{bz06}.  It appears that the confinement and the interior angular velocity depend sensitively on the boundary conditions imposed \citep{pg07}.  It remains unclear if a magnetic field can effectively render uniform rotation in the radiative interior.

The hydrodynamic model utilizes gravity waves to extract angular momentum from the solar radiative interior.  Internal gravity waves (IGW) are thought to be excited in the stable interior by convective downwellings hitting the convective-radiative interface.  Such waves can cause long range angular momentum redistribution.  Whereas earlier models failed to recognize the ``anti-diffusive" nature of gravity waves \citep{kq97,ztm97}, more recent models incorporate this behaviour into the theory \citep{ktz99,tkz02}, which can be briefly explained as follows.  Internal gravity waves generated at the convective-radiative interface propagate into the solar tachocline and set up shear layer oscillations (SLO), similar in nature to the Plumb \& McEwan experiment \citep{pm78}. This time-dependent, depth-dependent (but spherically-averaged) mean zonal flow is assumed to have a slightly stronger prograde sense than retrograde (on average), a consequence of the continual spin down of the Sun's convection zone by the magnetic torque arising from the solar wind outflow.  Because of the Doppler effect, prograde waves are preferentially dissipated in the upper part of the radiative region, allowing predominantly retrograde waves (transporting negative angular momentum) to propagate into the deep radiative interior.  It has been proposed that the solar radiative interior is thus spun down, tending ultimately toward a state of near uniform rotation \citep{tkz02}.  

Such wave driven shear flow oscillations have been observed in laboratory experiments \citep{pm78}, in the Earth's atmosphere \citep{ba01} and in Jupiter's atmosphere \citep{leo91}.  In addition, there is some evidence for oscillatory flows (the source and persistence of which is unknown) in the solar tachocline \citep{ho00}.  Therefore, it is not unreasonable to postulate such flows in the solar radiative interior.  There are, however, several issues with this model.  First, it is unclear that such oscillations could develop, at least in the upper part of the tachocline, due to the constant bombardment by overshooting plumes which likely disrupts the wave-mean flow interaction.  Second, the development of oscillatory flow depends sensitively on the wave spectrum and amplitudes assumed, which is, at best, poorly understood.  In fact, the model described above completely neglects the wave spectrum that would be produced by overshooting plumes.  Furthermore, the model does not address the effect of a latitudinally varying differential rotation on the wave generation, which may play a large role in the nature of waves which propagate into the deep interior \citep{fritts98}.  Finally, even if oscillations could develop in the simplified sense described, it is unclear how Doppler filtering could produce ${\it uniform}$ rotation in the deep radiative interior.  Even if only retrograde waves are allowed passage into the deep interior, as soon as one of them dissipates a radial differential rotation is established.  Subsequent waves propagating into this differentially rotating region will be Doppler shifted and the radial differential rotation enhanced.  
 
The study of internal gravity waves in astrophysical settings can take advantage of the vast literature on the subject in the atmospheric and oceanic communities.  Much of the research on angular momentum transport by IGW was spurred by the observation of the Quasi-Biennial Oscillation (QBO), which manifests itself as alternating easterly and westerly winds in the equatorial stratosphere.   While the QBO was first observed in 1960 \citep{reed61,eb61}, the first satisfactory theoretical explanation awaited \cite{lh68} and \cite{hl72}.  While different flavors of this original theory exist, the fundamental theoretical explanation has persisted and has been borne out in experiments such as the remarkable Plumb \& McEwan (1978) experiment.   However, despite the recognition of the physical mechanism which drives the QBO and significant observational details, simulating it from first principles has been challenging.  The spectrum of waves needed to force the oscillation is still a topic of research \citep{ba01,du97}.  Not all Global Climate Models (GCM) are able to produce a realistic QBO; these simulations depend sensitively on the convective parametrization, resolution and diffusivities \citep{ba01}.  So, whereas the basic properties and nature of the QBO in the Earth's atmosphere are understood, the details are not. 

In general, most theoretical studies of wave driven mean flows have been restricted to a quasi-linear formulation, a perturbation approach that accounts for the effects of weakly non-linear waves on the mean flow dynamics \citep{lh68,hl72,du97,ktz99,km01}. Clearly, this approach limits the wave dynamics and breadth of solutions that can be obtained.  The fully nonlinear numerical simulations described in the present paper, do not employ the approximations and parameterizations of the quasi-linear models in order to better understand wave driven mean flows and, in particular, those interactions under solar-like conditions.  These numerical simulations are conducted in an attempt to shed light on how such waves, and the mean flows they produce, can contribute to the rotational profile of the solar radiative interior.  

The remainder of the paper is organized as follows.  In section 2, we describe our numerical technique and in section 3, we briefly discuss the basics of wave-mean flow interaction.  In section 4, we present our results.  We discuss the major conclusions and their implications in section 5.

\section[]{Numerical Model}
\subsection{Equations}

The Navier-Stokes equations are solved for a perfect gas within the anelastic approximation.  Our model is two dimensional (2D), representing only the equatorial plane.  The anelastic approximation is appropriate when the flow speed is much smaller than the sound speed in the medium and when thermodynamic perturbations are small compared to the reference state; both of these constraints are well satisfied in the Sun's radiative interior.  However, unlike the Boussinesq approximation, the anelastic approximation allows a stratification in the reference state density and temperature.  The following equations are solved for the fluid flow relative to the rotating frame of reference and thermodynamic perturbations relative to a hydrostatic reference state:

\begin{equation}
\nabla \cdot \overline{\rho} \vec{v} = 0 
\end{equation}

\begin{eqnarray}
\lefteqn{\dxdy{\vec{v}}{t}+(\vec{v}\cdot\nabla)\vec{v}=-\nabla P - C\overline{g}\hat{r} + 2(\vec{v}\times\Omega)+}\nonumber\\
&    &\overline\nu(\nabla^{2}\vec{v}+\frac{1}{3}\nabla(\nabla\cdot\vec{v}))
\end{eqnarray}

\begin{eqnarray}
\lefteqn{\dxdy{T}{t}+(\vec{v}\cdot\nabla){T}=-v_{r}(\frac{d\overline{T}}{dr}-(\gamma-1)\overline{T}h_{\rho})+}\nonumber\\&  
& {(\gamma-1)Th_{\rho}v_{r}+\gamma\overline{\kappa}[\nabla^{2}T+(h_{\rho}+h_{\kappa})\dxdy{T}{r}]}
\end{eqnarray}

In these equations overbars represent reference state variables which are functions of radius only; the variables lacking overbars are the perturbations, which are functions of time, radius ($r$) and longitude ($\phi$).  We assume no flow or gradients in latitude.  Equation (1) is the anelastic version of the continuity equation.  In equation (2), the momentum equation, $\vec{v}$ is fluid velocity (with components $v_{\phi}$ and $v_{r}$), $P$ is the reduced pressure (equal to the pressure perturbation $p$ divided by the reference state density $\overline{\rho}$), $C$ is the codensity $(T/\overline{T} +p/(\overline{g}\overline{\rho}\overline{T})d\overline{T}/dr)$ \citep{bra95,rg06}, $T$ is the temperature perturbation, $\overline{g}$ is gravitational acceleration and $\overline{\nu}$ is the viscous diffusivity.  For the models presented here, $\Omega$, the angular velocity of the rotating frame, is a constant and in the $\hat{z}$ direction\footnote{Models with differential rotation imposed have been run, but for any reasonable differential rotation there is no effect on the wave dynamics because $\delta\Omega$ is much smaller than the wave frequencies explored.}.

Equation (3) is the heat equation written in terms of temperature.  The first term on the right hand side is the radial velocity times the prescribed super- or subadiabaticity, dictating the convective stability as a function of radius.  $d{\overline{T}}/dr$ is the reference state temperature gradient and $(\gamma-1)\overline{T}h_{\rho}$ is the adiabatic temperature gradient for a perfect gas, where $\gamma$ is the ratio of specific heats $c_{p}/c_{v}$ and $h_{\rho}$ is the inverse density scale height, $d \rm{ln} \rho / dr$.  The thermodynamic diffusivity, $\overline{\kappa}$, is depth dependent and $h_{\kappa}$ is $d \rm{ln} \overline{\kappa}/dr$.
 
For numerical convenience we solve equations (1) and (2) using a vorticity-streamfunction formulation.  Taking the curl of the momentum equation and imposing mass conservation we get the vorticity equation:

\begin{equation}
\dxdy{\omega}{t}+(\vec{v}\cdot\vec{\nabla})\omega=(2\Omega + \omega)h_{\rho}v_{r}-\frac{\overline{g}}{\overline{T}r}\dxdy{T}{\phi}-\frac{1}{\overline{\rho}\overline{T}r}\frac{d\overline{T}}{dr}\dxdy{{\it{p}}}{\phi} +\overline{\nu}\nabla^{2}\omega
\end{equation}
where $\omega$ is the $\hat{z}$ component of vorticity, $\nabla\times\vec{v}$.  For simplicity we have neglected the additional viscous terms due to the radial gradient of the viscous diffusivity.  Equation (4) is solved together with an equation relating the streamfunction, $\psi$, defined as $\overline{\rho}\vec{v}=\nabla\times\psi\hat{z}$, and the vorticity:

\begin{equation}
-\omega\overline{\rho}=\dxdys{\psi}{r}+(\frac{1}{r}-h_{\rho})\dxdy{\psi}{r}+\frac{1}{r^2}\dxdys{\psi}{\phi}
\end{equation}

This formulation automatically enforces equation (1). The independent variables directly calculated are then the temperature perturbation, T, the vorticity, $\omega$, and the streamfunction, $\psi$.   

\subsection{Numerical Technique}

Equations (1)-(3) describe internal gravity wave dynamics in the stable solar interior, neglecting the influence of a magnetic field.  The equations are solved in two dimensions in cylindrical coordinates representing the equatorial plane of the Sun.  The computational domain extends from $0.01R_{\odot}$ to $0.71R_{\odot}$, representing only the Sun's stable radiative interior.  The radially dependent reference state variables (density, temperature, subadiabaticity and gravity) are taken from a 21st-order polynomial fit to the one dimensional standard solar model Model S \citep{jcd91}.  The equations are solved using a Fourier decomposition in the longitudinal direction and a finite difference method in the radial direction.  The solutions are time advanced using a second order Adams Bashforth method for the nonlinear terms and an implicit Crank-Nicolson method for all the linear terms.  A spectral transform method is employed to compute the nonlinear terms in grid space each time step.  With a few exceptions, the majority of the models have a spatial resolution of 1000 (nonuniform) radial grid points x 512 longitudinal grid points.  The model is parallelized using Message Passing Interface (MPI).  Boundary conditions imposed on the bottom boundary of the domain are stress-free, impermeable and isothermal.  The conditions on the top are stress free and isothermal.  Waves are driven at the top boundary via a prescribed time-dependent permeable top boundary condition.

\subsection{Wave Driving}

More realistic models of the solar radiative interior also simulate the dynamics of the convection zone \citep{rg06}.  However, in these models the spectrum of motion in the radiative interior is complicated because of the turbulent convective overshoot, making a sensitivity study of fundamental wave dynamics virtually impossible.  Therefore, unlike previous models, here we exclude the convection zone and drive the gravity waves artificially at the top boundary of the model, which represents to the top of the radiative region.  The simplest way to produce a wave driven mean flow is to excite a prograde wave and a retrograde wave with the same frequency, wavenumber and amplitude.  These waves would combine to form a standing wave in the model's rotating frame of reference if there were no mean flow relative to this frame.

We employ the following procedure. 
The independent variables, vorticity, streamfunction and temperature, are expanded in a Fourier series in longitude, $\phi$, which is in radians.

\begin{equation}
\omega(r,\phi,t)=\sum_{m=0}^M(\omega^{c}_m(r,t)\cos(m\phi)+\omega^{s}_m(r,t)\sin(m\phi))
\end{equation}
\begin{equation}
\psi(r,\phi,t)=\sum_{m=0}^M(\psi^{c}_m(r,t)\cos(m\phi)+\psi^{s}_m(r,t)\sin(m\phi))
\end{equation}
\begin{equation}
T(r,\phi,t)=\sum_{m=0}^M(T^{c}_m(r,t)\cos(m\phi)+T^{s}_m(r,t)\sin(m\phi))
\end{equation}
where $m$ is the horizontal wavenumber.  Separate equations are solved for the sine and cosine coefficients of these variables.

A wave is excited at the top boundary as a radial velocity by prescribing a longitudinal and time dependent streamfunction there.  Most models are run with amplitude constant in {\it spectral} space.  Thus, equal amplitude means the waves have an equivalent amplitude $\psi$.  This means that the waves have slightly different amplitudes in grid space (since $v_{r} \approx m\psi/\overline{\rho} r$).  Variance of amplitude in grid and spectral space will be discussed further in section 4.  We further spectrally decompose $\psi$ in frequency space
\begin{eqnarray}
\psi(r,\phi,t)=\sum_{n=0}^N\sum_{m=0}^M(\psi^{cc}_{m,n}(r)\cos(m\phi)\cos(n \sigma_o t)\nonumber\\ +\psi^{cs}_{m,n}(r)\cos(m\phi)\sin(n \sigma_o t)\nonumber\\ +\psi^{sc}_{m,n}(r)\sin(m\phi)\cos(n \sigma_o t)\nonumber\\ +\psi^{ss}_{m,n}(r)\sin(m\phi)\sin(n \sigma_o t))
\end{eqnarray}
Here, $\sigma_o$ is the lowest frequency (in radians/second) considered.  After some algebra (D.O. Gough, private communication) one can show that, for a positive $m$, the amplitudes of a prograde wave $(m\phi - n \sigma_o t)$ and a retrograde wave $(m\phi + n \sigma_o t)$ are, respectively,
\begin{equation}
P=\frac{1}{2}[(\psi^{cc}_{m,n}+\psi^{ss}_{m,n})^2+(\psi^{cs}_{m,n}-\psi^{sc}_{m,n})^2]
\end{equation}
\begin{equation}
R=\frac{1}{2}[(\psi^{cc}_{m,n}-\psi^{ss}_{m,n})^2+(\psi^{cs}_{m,n}+\psi^{sc}_{m,n})^2] .
\end{equation}

The simplest way to force a wave driven oscillation is to force prograde and retrograde waves with the same amplitude.  By inspection of equations (10)-(11) one can see that a way to do this would be to force only one component of the four $\psi^{cc}_{m,n},\psi^{cs}_{m,n},\psi^{sc}_{m,n},\psi^{ss}_{m,n}$\footnote{Clearly other combinations could work, but this is the simplest.}; we choose $\psi^{sc}_{m,n}$.  That is, we force one or more wavenumber modes $\psi^{s}_m(r,t)$ on the top boundary with one or more frequency modes $\cos(n \sigma_o t)$ relative to the rotating frame freference.

\section{Fundamentals of wave driven shear flows}\label{sec:fundamentals}
The dynamics of wave driven mean flows has been studied extensively \citep{ho04,li90,plumb77}.  Here we review some of the basics.  When a single propagating wave\footnote{By ``wave" we mean a non-axisymmetric oscillating flow represented by a single spectral wave number, $m$, not equal to zero; whereas a ``mean flow" is an axisymmetric zonal flow with a spectral wave number of zero.} is attenuated it transfers angular momentum to the mean flow \citep{ep60}.  The forcing of the mean flow by waves depends on the wave attenuation, which depends on the frequency of the waves relative to the mean flow, which in turn depends on the shear in the mean flow.  Therefore, wave-mean flow dynamics is highly nonlinear.

One can appreciate how angular momentum is transferred from waves to the mean flow by considering the longitudinally-averaged horizontal component of the momentum equation for a simple constant density fluid \citep{ho04,li90}. 
\begin{equation}
\dxdy{\overline{U}}{t}+\frac{1}{r}\dxdy{~r<u'w'>}{r}=\nu(\dxdys{\overline{U}}{r}+\frac{1}{r}\dxdy{\overline{U}}{r})
\end{equation}
where u' is the azimuthal velocity, $v_{\phi}$, and w' is the radial velocity, $v_{r}$.  This equation shows that the mean zonal flow, $\overline{U}$, is driven by the divergence of the horizontally-averaged Reynolds stress (HARS) and is smoothed by viscous dissipation.  A simple way to understand this transport is to consider the product of two waves, one with wavenumber +m, the other with wavenumber -m.  This product equals the sum of two waves, one with wavenumber 2m and the other with wavenumber 0, which contributes to the mean flow.  

The mean zonal flow maintained by internal gravity waves is typically depth dependent, i.e., a ${\it shear}$ flow.  This "anti-diffusive" nature of gravity waves can be explained heuristically.  Imagine a prograde (m$>$0) wave and a retrograde wave (m$<$0) excited at the same radius.  If the angular velocity, $\Omega(r)$, of the medium increases inward, the prograde wave is Doppler shifted to lower frequency as it spirals inward ({\it measured from the mean flow}), whereas the retrograde wave is shifted to higher frequency.  These frequency shifts are relative to the frequency at which they were generated, $\sigma_{gen}$, where the angular velocity is $\Omega_{gen}$.  This Doppler shifted frequency is
\begin{equation}
\sigma(r)=\sigma_{gen}+m[\Omega_{gen}-\Omega(r)] .
\end{equation}
Since the dissipation rate, and therefore damping distance, are strongly frequency dependent, the prograde and retrograde waves dissipate at different depths.  A prograde wave transports positive angular momentum, whereas a retrograde wave transports negative angular momentum.  Therefore, where a prograde wave is dissipated the mean zonal flow is accelerated and where a retrograde wave is dissipated the mean flow is decelerated.  In this way two waves excited at a given radius with the same frequency and wavenumber but spiralling in opposite directions inward can modify the radial gradient of angular velocity, causing it to migrate upward toward the source of the waves and periodically reverse the direction of the radial shear.  \cite{plumb77} showed that two waves are unstable and will produce a shear, even in the absence of an initial shear.

\subsection{Quasi-Linear Approach}
To investigate wave-driven flows in detail the full set of fluid equations should be solved explicitly, so that the HARS appearing in (12) can be calculated directly.  However, this is computationally demanding as it requires good spatial resolution for both the waves (small spatial scales) and the mean flows (large scales) and good temporal resolution for the waves (short time scales) and the mean flow (long scales).\footnote{For example, the QBO has a period of approximately 28 months, but is thought to be driven by waves with periods less than approximately three days \citep{ba01}}  To avoid these difficulties researchers employ a variety of approximations, whereby the horizontally averaged Reynolds stress of a wave is evaluated according to linear theory. 

In the quasi-linear formalism the equation for the mean flow, such as that seen in equation (12), is subtracted from the full fluid equation to yield equations for the disturbances such as u' and w'.  The "mean field" approximation is then employed \citep{he63} in which the combination $(u'w')_{r}-<u'w'>_{r}$ (where the subscript "r" denotes derivation with respect to radius) is assumed small compared to other terms in the equation.  This renders the equations {\it linearized} thus allowing the disturbances to be written as simple sinusoidal functions of space and time.  The horizontal average of these fluctuations is then identified as a wave momentum.  

This treatment is similar to that of Bretherton (1966) in which the wave energy equation can be reduced to a conservation equation for the wave action density:

\begin{equation}
\dxdy{A}{t}+\nabla\cdot{Av_{g}}=0.
\end{equation}
Here the action density A is equal to E/$\omega$*, where E is the wave energy and $\omega$* is the frequency in the inertial frame.\footnote{Note that there is a correspondence between the action flux and the Reynolds stress.}  The product of the wave action density and the vertical group velocity, $v_{g}$ yields the wave momentum flux, F.  This conservation equation is derived in the absence of dissipative mechanisms.  In order to use this conservation equation in describing mean flow evolution, dissipation has to be included.  This is done in an ad hoc manner, yielding an equation of the form:
\begin{equation}
\dxdy{A}{t}+\nabla\cdot{F}=-F/L
\end{equation}
in which L is meant to represent a damping length.  Exponential decay (in space) of the wave flux is then retrieved by neglecting $\partial{A}/\partial{t}$.  This omission requires that the wave action vary slowly, similar to the assumption of WKB theory.  This formalism results in the equation that is typically solved:

\begin{equation}
\dxdy{\overline{U}}{t}+\sum\frac{1}{r}\dxdy{~r F_{s}e^{-\tau}}{r}=\nu(\dxdys{\overline{U}}{r}+\frac{1}{r}\dxdy{\overline{U}}{r})
\end{equation}
where $\tau$ represents a damping optical depth \citep{ktz99}.  A sum has been included to account for the flux from multiple waves, since the formalism described above applies to individual, non-interacting wave packets.  Both of the approaches described above neglect wave-wave interactions.

In the treatment described above the fluid is "weakly turbulent", in the sense that the nonlinear terms involving wave self interactions and their influence on the mean flow are retained, but those involving wave-wave interactions are neglected.  This approximation is valid when the divergence of the Reynolds stress term, $(u'w')_{r}$ is small compared with the inertial term $\partial{u'}/\partial{t}$, this limit is often described by requiring a small Froude number (w'/NL), where w' represents a typical wave velocity, L represents a typical length scale and N represents the Brunt-Vaisala frequency.  Other nonlinear effects, such as critical levels (where $\sigma-mU/r$ approaches zero) and internal reflection (where $\sigma$ tends to N) are also not described in the above formulation \citep{booker67}.

\subsection{Attenuation Mechanisms}
Eliassen \& Palm (1960) showed that in the absence of damping, forcing or critical layers, no momentum could be transferred from the waves to the flow.  Stated another way, {\it divergence} of the Reynolds stress (12) is needed to maintain mean flows.  The primary ways in which waves can be attenuated are by radiative dissipation or by critical levels.  Internal gravity waves are, in essence, thermal perturbations and therefore, are subject to thermal diffusion.  As a wave propagates vertically, the wave's amplitude is decreased due to radiative diffusion.  This dissipation leads to a nonzero divergence of the Reynolds stress, hence leading to angular momentum transfer and acceleration of the mean flow (12).  The radiative dissipation of the wave depends on the wave properties, such as the frequency and the wavenumber.  One can define a radiative damping {\it distance} \citep{ktz99}
\begin{equation}
d=\frac{\sigma^{4}}{\kappa k^3N^{3}}
\end{equation}
where $\sigma$ is the wave frequency, $\kappa$ is the thermal diffusivity, $k$ is the wavenumber (m/r) and N is the Brunt-Vaisala frequency.  This equation says that the higher the wave frequency and the smaller the wavenumber the less rapidly the wave dissipates.  That is, the higher the frequency the less time there is to diffuse and the larger the wavelength the smaller the spatial gradients that drive diffusion.   Radiative dissipation occurs continually as the wave propagates; therefore the resulting mean flow acceleration is very gradual.

Another way in which waves are attenuated is by critical level absorption.  A critical level is defined as that radius where the mean zonal velocity, $\overline{U}$, equals the horizontal phase speed of the wave, $c_{ph}=\sigma/k$.  In reality a wave packet has a range of horizontal phase speeds; therefore the critical level is more appropriately referred to as a critical ${\it layer}$.  The inviscid WKB solution is singular at this radius.  \cite{booker67} studied the wave-critical level interaction relaxing the WKB approximation and showed that, in the limit of large Richardson number ($Ri=N^{2}/(dU/dz)^{2}$), the wave energy density is strongly attenuated at a critical level with nearly complete momentum transfer to the mean flow.  More recent studies \citep{wda89,wda94} indicate that some wave energy is transmitted through the critical level and some is reflected, but approximately one-third of the wave energy accelerates the mean flow.  Mean flow acceleration at a critical layer is thus rapid and relatively local.  

\section{Results}

\subsection{Two Wave driven shear oscillations}
\subsubsection{Model Dependencies}
We simulate a mean flow oscillation maintained by two waves excited at our top boundary: a prograde and retrograde wave each with the same longitudinal wavenumber and frequency.  There is no background rotation in this case.  Figure 1 shows the resulting mean angular velocity as a function of time and radius for model M2 (see Table 1).  Red and blue represent prograde and retrograde mean angular velocity, respectively.  This figure illustrates the salient features of a gravity wave driven oscillating shear flow: (1) periodicity and persistence, (2) propagation of the mean flow pattern toward the wave source in time and (3) a doubly peaked shear layer at all times.  All of the oscillating models listed in Table 1 exhibit this fundamental behaviour.  

Figure 2 shows snapshots of the flow evolution during one cycle. Figures 2a and 2b show the full disk evolution of vorticity (m$\neq$0), whereas (c) and (d) show the mean angular velocity as a function of radius.  One can see both the mean flow and the wave dynamics in this figure.  In (c), the secondary mean flow (i.e., the deeper one) is prograde and in (a) the waves in this secondary layer are propagating in the ${\it retrograde}$ (i.e, clockwise) direction. Note that since these waves are driven at the top boundary the energy, which is transported by the group velocity, spirals inward.  Therefore, since the phase velocity of internal gravity waves is perpendicular to the group velocity, the phase velocity has a dominantly outward component in addition to its prograde (i.e, counter-clockwise) or retrograde (i.e., clockwise) tilt.  This phase velocity is easily seen in movies of these simulations.  In the snapshot illustrated in Figure 2a the phase is propagating upward with a retrograde tilt in the secondary, whereas in the primary (the upper layer) the phase is propagating upward with a prograde tilt.  The opposite configuration is occurring in Figure 2b, which is later in time.

The existence of predominantly retrograde waves where the mean flow is prograde occurs because the prograde waves have been dissipated in this region; thus giving rise to the prograde mean flow.  This observation also means that some of the energy in the retrograde waves is able to propagate through the retrograde primary in Figure 2a (this is discussed further in 4.1.3).

The mean flow oscillation depends on the wave properties (wavenumber, frequency and amplitude), as well as the properties of the medium in which the waves propagate.  A series of models were run, varying these parameters to investigate the various dependencies.{\bf\footnote{Note again, that the amplitude is in spectral space, $v_{r} \approx m\psi/r\overline{\rho}$.}}.  Table 1 lists the various two-wave models and their parameters.\footnote{The models presented here represent a small subset of the models attempted.  This subset represents those values of the various parameters that could be reasonably resolved in a timely manner.}  These dependencies will be discussed only briefly here since they have been reported previously \citep{plumb77,wedi06}.

\begin{table*}
\centering
\begin{minipage}{150mm}
\caption{Model parameters. ${\it m}$ is the horizontal wavenumber, $f$ is the wave frequency ($\sigma/(2\pi)$) and A is the forced amplitude of the streamfunction, $\psi$, at the top boundary, $\kappa$ and $\nu$ are the thermal and viscous diffusivities, D is the depth of oscillation (the depth is measured as that point were the mean angular velocity drops significantly below $10^{-7}$ rad/s) and P is the period of the oscillation in days.  AV and CP represent the maximum angular velocity of the mean flow and horizontal phase speed of the driven wave ($\sigma/m=2mf\pi$), respectively.  The asterisk on the angular velocities for model M10 and M16 indicates that the pro- and retrograde peak angular velocities were ${\it not}$ equal as in the rest of the models.  The value listed is an average.}

\begin{tabular}{llllllllll}

\hline
Model & m & f($\mu$ Hz) & A($10^{13} gm/(cm s)$) & $\kappa$($10^{11}cm^2/s$) & $\nu$($10^{11}cm^2/s$) & D($10^{9}cm$) & P($days$) & AV($10^{-6}rad/s$) & CP($10^{-6}rad/s$) \\
\hline
M1 & 10 & 10 & 1 & 1 & 1 & 8.8 & 544.0 & 6.30 & 6.28\\
M2 & 15 & 10 & 1 &1 & 1 & 4.4 & 86.80 & 4.28 & 4.18\\
M3 & 20 & 10 & 1 &1 & 1 & 2.9 & 25.40 & 3.18 & 3.14\\
M4 & 30 & 10 & 1 &1 & 1 & 2.0 & 2.900 & 0.65 & 2.09\\
M5 & 15 & 8 & 1 &1 & 1 & 3.1 & 28.90 & 3.33 & 3.35\\
M6 & 15 & 15 & 1 &1 & 1 & 10.2 & 463.0 & 6.44 & 6.28\\
M7 & 15 & 20 & 1 &1& 1 & 17.5 & 1620. & 8.53 & 8.37\\
M8 & 15 & 10 &1& 1 & 0.2 & 4.9 & 144.7 & 5.20 & 4.18\\
M9 & 15 & 10 & 1& 1 & 0.5 & 4.9 & 119.2 & 4.70 & 4.18\\
M10 & 15 & 10 & 1& 1 & 5.0 & NA & steady & 1.30* & 4.18\\
M11 & 15 & 10 & 1& 0.1 & 1 & 6.3 & 189.8 & 4.20 & 4.18\\
M12 & 15 & 10 & 1& 0.5 & 1 & 5.4 & 113.4& 4.30 & 4.18\\
M13 & 15 & 10 & 1& 3.0 & 1 & 2.9 & 50.90 & 3.90 & 4.18\\
M14 & 15 & 10 & 2& 1 & 1 & 4.39 & 26.60 & 4.85 & 4.18\\
M15 & 15 & 10 & 0.5& 1 & 1 &  5.36& 405.0 & 3.75 & 4.18\\
M16 & 15 & 10 & 0.2& 1 & 1 & NA & steady & 0.14* & 4.18\\
\hline
\end{tabular}
\end{minipage}
\end{table*}

As is discussed in 4.1.2, we find that the mean flow oscillation period is determined largely (although not entirely) by radiative dissipation.  Therefore, properties of the oscillation such as the period and depth can be qualitatively understood in terms of the radiative dissipation of waves; waves which are dissipated more rapidly produce shallower mean flow oscillations with shorter periods.  This can be understood in terms of the gravity wave phase speed: waves with higher phase speed propagate further in a diffusion time.  Therefore, gravity waves with larger frequency or smaller wavenumber produce mean flow oscillations that penetrate deeper and have longer periods.  Likewise, larger thermal and viscous diffusivities lead to mean flow oscillations with shorter periods and shallower depths.  Whereas the energy dissipation of the waves is predominantly thermal, the resulting angular momentum transfer is mediated through viscosity (12); therefore viscous diffusivity qualitatively affects the oscillation period and depth in a way similar to that of thermal diffusivity.

Using the simplified quasi-linear approach discussed in Section 3, \cite{km01} show that the nature of the mean flow oscillations  produced (beyond period and depth) depends sensitively on the viscous dissipation.  They find that as the viscous diffusivity is decreased, the mean flows progress from steady shear (no oscillation), to periodic oscillations, to chaotic reversals.  

The results presented here generally agree with those of \cite{km01} with one important exception.  If the viscous diffusivity of model M2 is increased to $5\times 10^{11} cm^{2}s^{-1}$ (a factor of five), keeping other parameters fixed, the mean flow reverts to a steady shear flow.  As the viscous diffusivity is decreased (M8-M10, displayed in Table 1) periodic reversals develop and the peak mean angular velocity increases.  However so far we have not been able to produce chaotic solutions in these simple two-wave models.  Our simulations with viscous diffusivity less than $10^{10} cm^{2}s^{-1}$ are not well resolved (keeping other parameters fixed).  The same behaviour can be recovered, in the opposite sense, by varying the amplitude of the forcing (Table 1, Models M14-M16).  As the forcing amplitude is increased flows progress from steady to periodic oscillations and the peak mean angular velocity increases.  Note, we have run cases with much smaller driving amplitudes (down by 100) {\it and} diffusivities (each down by 100) than those listed in Table 1. However, in those models a mean flow does develop slowly, but after 200 years of simulated time they have not reversed and has reached only half the critical amplitude.
\subsubsection{Mean Flow Evolution}
The major difference in the resulting behaviour in these simulations compared to those of \cite{km01} is the presence of critical layers.  As described in section 3, critical layers cannot occur in a quasi-linear formulation such as that used in \cite{km01}.  In the following discussion of critical layers we continue to refer to the ``primary'' layer as the mean flow that is closest to the wave source (the top in this case)  and the ```secondary'' as the mean flow below the primary; this is illustrated in Figure 3b.  We also refer to mean flows as being ``critical'' when the peak angular velocity meets or exceeds the horizontal phase speed of the driven wave and ``sub-critical'' when it does not.  

Figures 3 (a-e) show snapshots of the mean angular velocity during one cycle of model M12.  The dashed lines represent the horizontal phase speed of the driven wave relative to the rotating frame of reference; angular velocities with amplitudes greater than this are ``critical''.   After a reversal of the mean flow, the primary shear layer maintains its maximum amplitude while the secondary shear layer slowly grows and moves upward toward the source.  For example, when the primary layer is retrograde, as in Figures 3 (a-c), retrograde waves, observed relative to the mean retrograde flow at the top boundary, have a lower frequency and prograde waves have a higher frequency (equation 13).  As depth increases, the retrograde flow decreases and becomes prograde within the secondary layer.  This favors the transmission of the retrograde waves and the radiative dissipation of the prograde waves, which makes the amplitude of the prograde secondary mean flow grow.  As prograde waves continually dissipate the shear, and thus frequency shift, becomes stronger, causing the peak of the secondary flow to move toward the source in time.  The growth of the secondary layer is relatively slow and takes approximately 85\% of the total oscillation period.  When the secondary layer is accelerated to an angular velocity approaching ``critical'' the prograde waves are rapidly dissipated, transferring the bulk of their angular momentum to the mean flow {\it above} the critical layer and bringing about the reversal of the primary.  The reversal takes only about 15\% of the total oscillation period.  The rapid acceleration associated with a critical layer begins before the secondary flow reaches a critical value.  This happens for two reasons.  First, as mentioned above, a wave ${\it packet}$ has a range of horizontal phase speeds and therefore there is a critical ${\it layer}$.  Second, there are regions in which the local (nonaxisymmetric) longitudinal velocity is critical, despite the mean flow being sub-critical.

Rapid dissipation above the critical layer and the associated angular momentum transfer is predicted based on the theories described in 3.2.  In our simulations we see that this rapid dissipation is mediated by nonlinear wave-wave interactions which transport wave energy from the driven wavenumber to higher harmonics.  This transport can be seen in Figure 4, which shows the wave energy as a function of radius and wavenumber during growth of the secondary shear layer (a) and during reversal of the primary (b), for model M9.  During the bulk of the cycle most of the energy is in the driven mode (m=15), with minimal energy in higher modes.  However, once the angular velocity approaches the critical value energy is transferred from the driven wave to higher harmonics where it can be more easily dissipated (because of the larger spatial gradients), thus bringing about rapid reversal of the primary.  Note, because of the quadratic nonlinearity, the energy is transferred only to horizontal modes that are multiples of the driving mode (m=15).

The quasi-linear description of wave driven mean flow oscillations omits wave-wave interactions and relies on viscous diffusion of the mean flow for the reversal. That is, the right hand side (rhs) of (12) must meet or exceed the wave forcing (second term on the left hand side (lhs) of (12)).  Figure 5 shows that this is not the case in our simulations.  Even during a reversal (Figure 5c), the viscous force acting on the mean flow is lower in amplitude than the divergence of the nonlinear Reynolds stress.  This is not to say that viscous dissipation does not play a role.   Viscosity, in addition to thermal diffusion, dissipates the waves and it acts on the mean flow.  However, the nonlinear wave-wave momentum transfer discussed above, followed by the dissipation of {\it waves} dominates the dynamics of the mean flow, not the dissipation of the mean flow.
  
Therefore, the presence of critical layers makes these models qualitatively different than those described by the classical quasi-linear theory.  This qualitative difference may produce quantitative differences in the depth and period of the oscillation.  While we have attempted models with significantly reduced amplitudes and diffusivities, we have not been able to produce an oscillation in $5\times 10^{9} s$ of integration time.  However, a shear flow with an amplitude approximately half the critical value has developed and is still growing, but the growth rate is very slow and we have concluded that resolving an oscillation under  such conditions is not numerically reasonable.

\subsubsection{The IGW Critical layer Interaction}

As discussed in 3.2, it has been demonstrated in several previous studies that when a gravity wave is incident on a critical layer, rapid dissipation and subsequent angular momentum deposition occurs, resulting in the acceleration of the mean flow.  We have discussed this process above as the mechanism that brings about rapid reversal of the primary shear flow.  However, Figure 3 (a-c) clearly shows that although the primary mean flow is indeed critical to the driven wave, the mean flow is not accelerated, despite the constant bombardment of waves.  Furthermore, as shown in Figure 2, despite a retrograde primary layer, retrograde waves are evident (indeed dominant) below in the secondary layer.  Clearly, waves are propagating past this apparent critical layer.

To investigate this somewhat surprising behaviour we spectrally decomposed the wave motion for model M2 in frequency space in order to distinguish between prograde and retrograde waves for model M2.  This was done during different times in the cycle representing either prograde or retrograde primary mean flows.  The results are shown in Figure 6.  When there is a prograde primary mean flow (top two plots), prograde wave energy is transferred to higher wavenumbers as discussed in 4.1.2.   However, the wave energy is additionally Doppler shifted to higher frequencies.  Because the shear is so strong, this Doppler shift is substantial, thus producing waves with ${\it higher}$ phase speeds despite the increased wavenumbers.  Waves with phase speeds higher than the phase speed of the driven wave (represented by the dashed line in Figure 6) will not recognize the mean flow as a critical level and therefore, propagate freely into the interior\footnote{The Doppler shift of waves at critical layers was recognized by \cite{fritts82} and termed "self-acceleration".}.  On the other hand, retrograde waves incident on the prograde primary do not experience a critical level, so there is very little energy found at higher harmonics.  The same behaviour (of the opposite sense) is observed when there is a retrograde primary (bottom two figures).  The net effect of an IGW-critical layer interaction depends crucially on the sense (prograde or retrograde) of both the waves and mean flow.  

As mentioned in 4.1.2, when the mean flow is sub-critical there are regions that have non-axisymmetric longitudinal velocities in excess of the horizontal phase speed and so are locally super-critical.  Likewise, when the mean flow is super-critical there are regions in which the local longitudinal velocity is sub-critical.  Given the complexity of natural systems, it is likely that such non-uniformity occurs and should be considered when discussing the propagation of waves near critical layers.  

As a wave approaches a critical level its vertical wavenumber is reduced substantially.  Therefore, numerical simulations of critical layers require fine radial resolution.  In order to test that the resolution employed was adequate to resolve the decrease in vertical wavenumber, we compared simulations with radial resolution 2.6 times finer and found that the same nonlinear transfer near a critical layer occurs.  Similarly, the period and depth of the mean flow oscillation were not affected by increased radial resolution.  Therefore, the flow behaviour in the vicinity of a critical layer should be adequately resolved.

\subsection{Forcing a Wave Spectrum}

The above results are observed when we force only one prograde wave and one retrograde wave, with the same amplitude, wavelength and frequency relative to the rotating frame.  Natural phenomena are clearly not as simple.  A time dependent spectrum of waves is excited in the solar interior and in the Earth's atmosphere.  To understand the generation of mean flow oscillations in nature we must understand how ${\it multiple}$ waves interact.
\begin{table*}
\centering
\begin{minipage}{150mm}
\caption{Models with multiple waves driven.  Columns are denoted as in Table 1.}

\begin{tabular}{llllll}

\hline
Model & m & f($\mu$ Hz) & A ($10^{13} gm/s)$& $\kappa$($10^{11}cm^2/s$) & $\nu$($10^{11}cm^2/s$) \\
\hline
MW1 & 10,15 & 10,15 & 0.5 & 1 &1  \\
MW2 & 5,10,15 & 5,10,15 & 0.5& 1 &1  \\
MW3 & 5,10,15,20 & 5,10,15,20 & 0.5&1 & 1\\
MC1 & 15 & 10 & 0.01 & 0.01 & 0.01 \\
MC2 & 10,15 & 10,15 & 0.01 & 0.01 & 0.01 \\
MC3 & 5,10,15 & 5,10,15 & 0.01 & 0.01 & 0.01\\
MC4 & 5,10,15 & 5,10,15 & 0.022 & 0.01 &0.01 \\
MSF5 & 5,10,15 & 5,10,15 & see text & 0.01 & 0.01 \\
MSF6 & 5,10,15 & 5,10,15 & see text & 0.01 & 0.01 \\
MSK5 & 5,10,15 & 5,10,15 & see text & 0.01 & 0.01\\
MSK6 & 5,10,15 & 5,10,15 & see text & 0.01 & 0.01 \\
MF1 & 5,10,15 & 5,10,15 & see text & 0.01 & 0.01 \\
MF2 & 5,10,15 & 5,10,15 & see text & 0.01 & 0.01 \\

\end{tabular}
\end{minipage}
\end{table*}

We conducted several experiments varying the number and make up (frequency, wavenumber and amplitude) of the driven waves.  These models are listed in Table 2.  In models MW1-MW3 we keep the amplitude of every wave fixed, but vary the number of waves forced.  For example, model MW1 forces wavenumbers 10 and 15, each with frequencies 10$\mu Hz$ and 15$\mu Hz$.  Therefore four prograde and four retrograde waves are forced, each with amplitudes the same as model M15.  If one considers wave energy to be $\propto v^{2}$ then models MW1, MW2 and MW3 have approximately 1.4, 1.5 and 3.5 times more energy input than M15 (considering the dependence of velocity on wavenumber).  We find that as we increase the number of waves, it is increasingly difficult to produce an oscillating shear flow.  MW1 produces an oscillation, but neither MW2 or MW3 result in oscillatory solutions.  In fact, the peak angular velocities (over the time run) for MW2 and MW3 are approximately 1000 and 100 times smaller, respectively, than MW1, despite having increased energy input.   

These smaller mean flows for MW2 and MW3 suggests that the forcing (second term on the left hand side of (12)) of the mean flow is lower, since the viscous diffusivity is held constant.  To investigate this we plot the HARS, $<u'w'>$, for models MW1-MW3 and M15 as a function of radius in Figure 7a.  There are a couple of interesting features to note.  First, the HARS does not increase with increased wave energy.  In fact, MW2 and MW3 have a lower HARS compared with model MW1 at the top of the domain.  Second, the radial profile of the HARS varies substantially.  The magnitude of the HARS for model M15 drops by nearly eight orders of magnitude over the radius range shown, whereas the multiple wave models drop by only two-three orders of magnitude.  This could be partially due to the wavenumbers and frequencies chosen; MW2 forces a lower frequency and wavenumber implying longer dissipation length by (17).  However, the difference is not entirely due to this effect as MW3 forces both lower and higher frequencies (than M15) and yet still has a significantly less steep slope.  The difference in slope has a drastic effect on the forcing of the mean flow since that forcing depends on the radial gradient of the HARS.  It is clear that the behaviour of the HARS for an ensemble of waves is significantly different from that of the sum of individual, non-interacting waves.

As discussed in Section 3, in order to approximate velocity fluctuations as simple sinusoidal functions of space and time the sum $\bf{(u'w')_{r}-<u'w'>_{r}}$ is assumed small in comparison to other terms in the equation. In Figure 7b, we show the ratio of the Reynolds stress to the HARS at a randomly chosen azimuth.  We see that when only one prograde and one retrograde wave are forced, $u'w'  \approx <u'w'>$ and the approximation employed in the quasi-linear approach could be valid.  However, as the number of waves is increased u'w' becomes significantly larger than $<u'w'>$.  In these cases in order to neglect the sum of these two quantities $(u'w')_{r}$ must be small in comparison to $\partial{u'}/\partial{t}$.   Stated another way, the Froude number for the wave (w'/Nz) \citep{am76}, where z is the vertical scale associated with a wave and the other quantities have their typical meanings, must be small.  Using velocity and vertical wavelength scales typical in figure 7b and a Brunt-Vaisala frequency of $10^{-4}$ one can estimate the Froude number to be approximately 0.1 indicating that nonlinear wave-wave interactions are not negligible.  The Froude number is however, just an estimate of the relative importance of the inertial term and the divergence of the Reynolds stress.  In these simulations we can directly measure this ratio, which we show in the following sections.

\subsubsection{Effect of Amplitude}
Nonlinear wave-wave interactions are not only important for large amplitude waves.  Models MC1-MC3 have been run with amplitudes and diffusivities each reduced by a factor of 100.  To compensate for the lower diffusivities we increased the numerical resolution by a factor of two in the horizontal direction and by a factor of 3.75 in the radial direction.  As for the models described above, when multiple waves are driven no oscillation develops in the 40 years simulated.  This is not entirely surprising, the low amplitude and diffusivities imply significantly longer oscillation periods (see Section 4.1.1).  We cannot realistically run simulations long enough to rule out oscillatory behaviour.  However, as in models MW2 and MW3, the nonlinear wave-wave interactions are non-negligible.  Figure 8a shows the radial gradient of the HARS (dashed lines) and the radial gradient of the Reynolds stress (solid lines) for models MC2 and MC3.  Figure 8b shows the radial velocity perturbation, w', as a function of time at a radius equivalent to $0.69R_{\odot}$ for the same models.  There are two main features to note.  First, the amplitude of the HARS is two orders of magnitude lower than the Reynolds stress (Figure 8a).  This just represents the fact that it is more difficult for low amplitude waves to force a mean flow, given the same integration time.  Therefore, in order to neglect the nonlinear term, $(u'w')_{r}$, one needs to show that this term is small compared to the inertial term, $\partial{w'}/\partial{t}$.  Upon inspection of Figure 8b one can easily estimate the amplitude of the inertial term as approximately $10^{-2}-10^{-3}$ depending on the model.  The Reynolds stress at this radius is also approximately $10^{-3}$; at best the Reynolds stress is 10\% of the inertial term, hardly negligible.  The amplitudes of these waves, depending on the wavenumber, varies between 50 cm/s and 150 cm/s.  These amplitudes combined with those modeled in MW1-MW3 represent any reasonable estimates for the wave amplitudes in the solar tachocline and radiative interior.

\subsubsection{Effect of Spectra}

The models listed above force waves with constant streamfunction, $\psi$, which makes the radial velocities proportional to m.  This gives a wave energy "flux" per mode proportional to $m^{3}$.  We ran five models in addition to models MC1-MC3 to investigate the dependence of the results on the wave spectrum.  These models are listed in Table 2 as models MC4 representing a model with wave flux proportional to $m^{3}$ but with an energy input larger than that of MC3 by a factor of five.  Models MSF5 and MSF6 represent models with a wave flux which is flat in {\it real} space (constant wave energy flux per mode), whereas models MSK5 and MSK6 represent models with a wave flux proportional to $m^{3}f^{-3}$, similar to that proposed by \cite{ktz99}.  Since the amplitudes of the individual waves in these models varies substantially, we do not list them in Table 2.  Furthermore, models appended with a "6" represent models with a total wave energy input five times larger than models appended with a "5".  The amplitudes of the velocities for models appended with "5" vary from 20 cm/s to 150 cm/s, whereas those appended with "6" vary from 60 cm/s to 340 cm/s.  These models were run for a total of 200 years.  Although, none showed oscillatory behaviour in that time, the models with higher input energy (MC4, MSF6, MSK6) do maintain strong shear flows.  Our results do not rule out oscillations with periods longer than hundreds of years.

The results for these models are shown in Figure 9 which represents the same physical quantities as portrayed in Figure 8: (a) and (b) represent the lower input energy models (MC3,MSF5,MSK5), while (c) and (d) show the various spectra models with higher input energy (MC4,MSF6,MSK6).  There are several interesting features.  First, the amplitude of the HARS, represented by dotted lines, is approximately four orders of magnitude smaller than the non-averaged Reynolds stress for the lower energy models.  Therefore, $(u'w')_{r}-<u'w'>_{r} \approx (u'w')_{r}$ and in order to neglect these terms $(u'w')_{r}$ must be small compared to $\partial w'/\partial t$.  Upon inspection of Figure 9b one sees the amplitude of the inertial term is approximately $4\times10^{-3}$ (this amplitude varies very little with spectrum); examining Figure 9a at the same radius (0.69$R_{\odot}$), one sees that the divergence of the Reynolds stress, $(u'w')_{r}$, has virtually the same amplitude.  Therefore, the Reynolds stress cannot be neglected and the problem is fundamentally nonlinear.  Finally, one can see that the slopes for models MSF5 and MSK5 (represented by blue and red, respectively) are significantly different than the slope for model MC3 (represented by black), despite having the same frequencies and wavenumbers driven.  The radiative wave damping for an individual wave, described by (17), implies that the slope should depend only on frequency, wavenumber, diffusivity and Brunt-Vaisala frequency.  These parameters do not change between models MC3, MSK5 and MSF5, yet the slopes change substantially.  This observation implies that not only are wave-wave interactions relevant but that they are spectrum dependent.  The higher input energy models MC4, MSK6 and MSF6 show virtually the same behaviour, with one exception.  In these models mean flows are more readily generated over the same period.  Whereas the HARS is significantly larger than in the lower energy models, it is still lower than the un-averaged Reynolds stress by one to two orders of magnitude.  One can make the same comparison between $(u'w')_{r}$ and $\partial {w'}/\partial {t}$ and find the same result; namely that the nonlinear terms are non-negligible.  

It is also worth pointing out that the wave flux predicted by \cite{ktz99}, represented by the red lines in Figure 9 appear to be more efficient at driving mean flows.  In both the low energy models and the high energy models the HARS is closer to the non-averaged Reynolds stress than in the other two models.  The wave-wave interactions appear to be spectrum dependent.  However, in addition, it appears that the (in)coherence of these interactions and their ability to force a mean flow is spectrum dependent.

\subsubsection{Effect of Initial Conditions}

All of the models described above start with no mean flow.  Therefore, if the waves cannot force a mean flow, by construct the problem is non-linear, in the sense that one cannot separate the flow into a mean (with large amplitude) and a wave (with smaller amplitude).  Therefore it may not be surprising that these models exhibit such nonlinear behaviour.  To investigate whether wave-wave interactions are influenced by the presence of a mean flow we ran two additional models, labeled in Table 2 as MF1 and MF2.   These models have a mean flow forced from the outset: MF1 has a sinusoidal function forced in the top $10^{9}$ cm (or 2\% of the radius of the solar interior), while MF2 has a half-sinusoid function in the top $10^{9}$ cm.  In both models the mean flow is specified to have a maximum amplitude of $10^{4} cm/s$ and is forced with a spectrum of waves which all contribute the same amplitude to $v_{r}$, making the total $v_{r}$ peak at 100 cm/s.  The mean flow is enforced such that wave-wave interactions and wave-mean flow interactions are allowed through the nonlinear terms.  
The results of these two models are shown in Figure 10; similar to figures 8 and 9 the upper panel shows the radial derivative of the Reynolds stress (solid lines), and the radial derivative of the HARS (dotted lines).  The Reynolds stress depicted is due to only wave-wave interactions, that is, wave interactions with the imposed mean flow is not included in the value shown in Figure 10.  Black represents model MF1, while blue represents MF2.  In these models the HARS is still an order of magnitude lower than the non-averaged Reynolds stress (Figure 10a), but again the Reynolds stress has an amplitude similar to $\partial w'/\partial t$ as measured in Figure 10b.  Even in models in which a mean flow is prescribed and the wave amplitude is prescribed to be 1\% of the mean flow amplitude, nonlinear interactions are non-negligible.

\section{Discussion}\label{sec:discussion}

We conducted highly simplified, but fully nonlinear, numerical experiments of internal gravity waves in the solar radiative interior to understand whether such waves can force oscillating shear flows.  Our results are threefold.  First, we produce oscillating shear flows for models in which a single horizontal wavelength is forced as equal amplitude prograde and retrograde propagating waves.  However, most of these highly simplified models exhibit critical layers which dominate the mean flow dynamics.  Second, when waves are incident on a critical layer the amount of energy deposited compared to the amount of energy transmitted varies depending on the sense of the mean flow those waves propagate into.  

Finally, when a spectrum of waves is forced the standard quasi linear approximation which neglects wave-wave interactions is inadequate.  In fact, none of the models we conducted could be adequately described by quasi-linear theory, either because of the presence of critical layers or because of the dominance of wave-wave interactions.  Such wave-wave interactions are non-negligible from waves ranging in amplitude from 50-1000 cm/s and diffusivities ranging from $10^{9}-10^{11} cm^{2}/s$ for various spectra of driven waves, with or without an enforced mean flow.  This range covers any reasonable estimate for flow velocities and turbulent diffusivities in the solar tachocline.  Lower diffusivities would make the nonlinear wave-wave interactions even more important.  Therefore, it appears that in order for quasi-linear theory to be valid for flows in the solar tachocline one would need unreasonably low amplitude waves combined with unreasonably high diffusivities.  

The ability for waves to enforce uniform rotation in the radiative interior depends sensitively on the ability to set up an oscillating shear layer in the solar tachocline.  It has been shown previously that that ability depends sensitively on the spectrum and amplitudes of the waves driven.  Here, we further show that the simplified theory used to describe those flows is inadequate for various spectra, amplitudes and diffusivities.

\bibliographystyle{mn2e}
\bibliography{igwcl}
\section*{Acknowledgments}

We thank M. McIntyre, D. Fritts and D. Gough for helpful discussions.  We also would like to thank our referee, J.P. Zahn for his insightful comments.  Support for this research was provided by a NASA grant NNG06GD44G.  T.R. is supported by an NSF Astronomy and Astrophysics Postdoctoral Fellowship under award 0602023.  Computing resources were provided by NAS at NASA Ames, by an NSF allocation at the Pittsburgh Supercomputing Center and by an NSF MRI grant AST-0079757.

\clearpage
\begin{figure}
\centering
\includegraphics[width=6in]{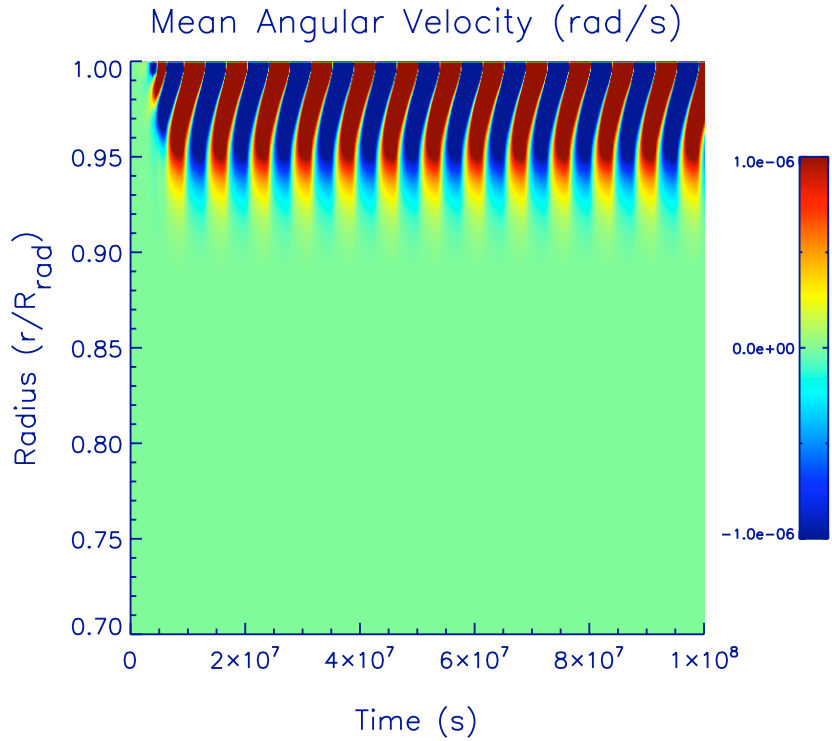}
\caption{Mean angular velocity as a function of radius and time for model M2.  Only the outer 30\% in radius of our simulated stable region is depicted here.  Prograde motion is represented by red, while retrograde motion is represented by blue.  The main features of a gravity wave driven mean flow oscillation are recovered.}
\end{figure}

\clearpage
\begin{figure}
\centering
\includegraphics[width=6in]{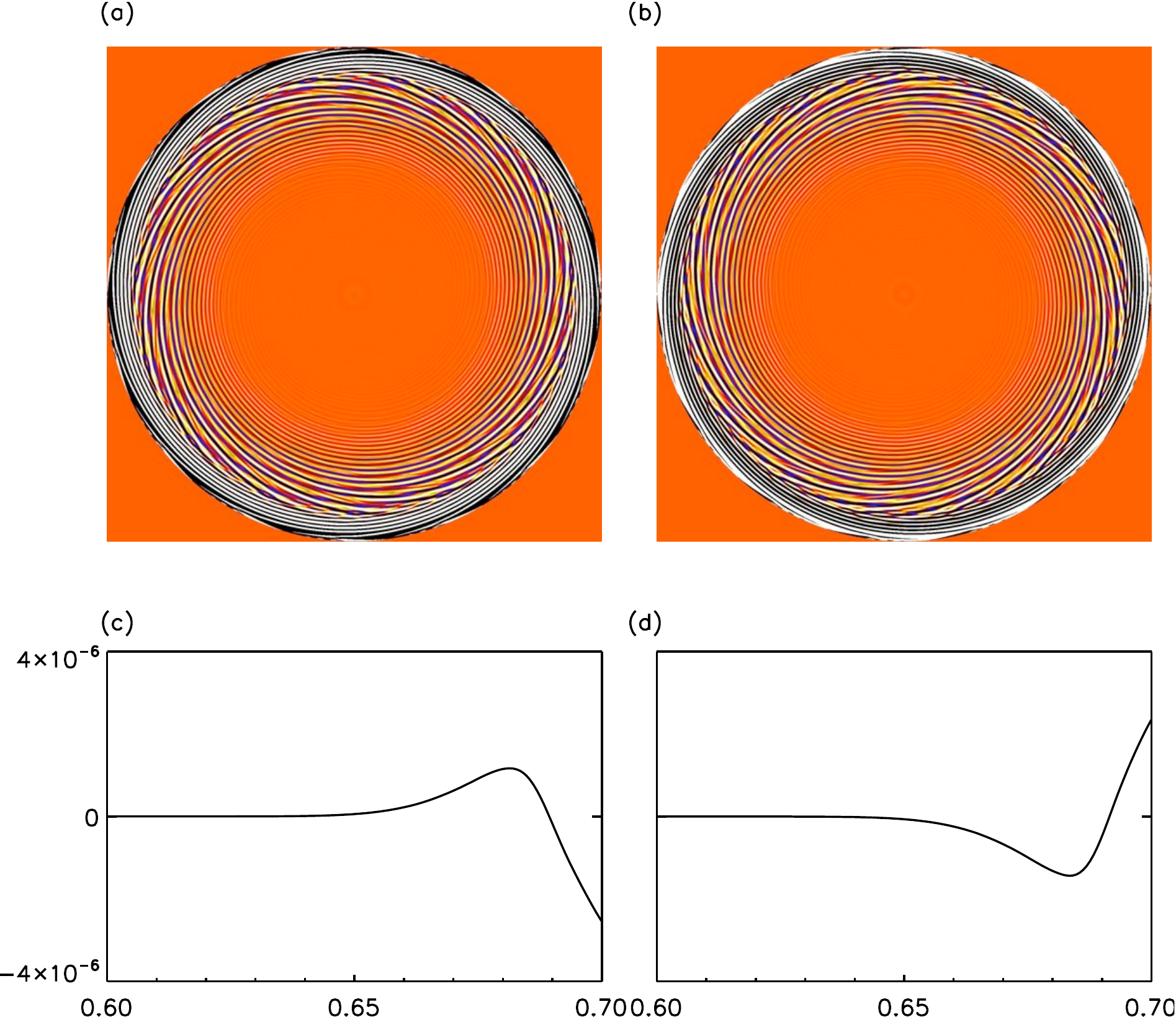}
\caption{Flow dynamics over a cycle for model M2.  Vorticity is depicted in (a) and at a later time in (b).  Positive vorticity is represented by yellow/white and negative vorticity represented by dark (blue/black) colors. (c) and (d) represent the mean angular velocity at the times corresponding to (a) and (b), respectively.}
\end{figure}
\clearpage
\begin{figure}
\includegraphics[width=6in]{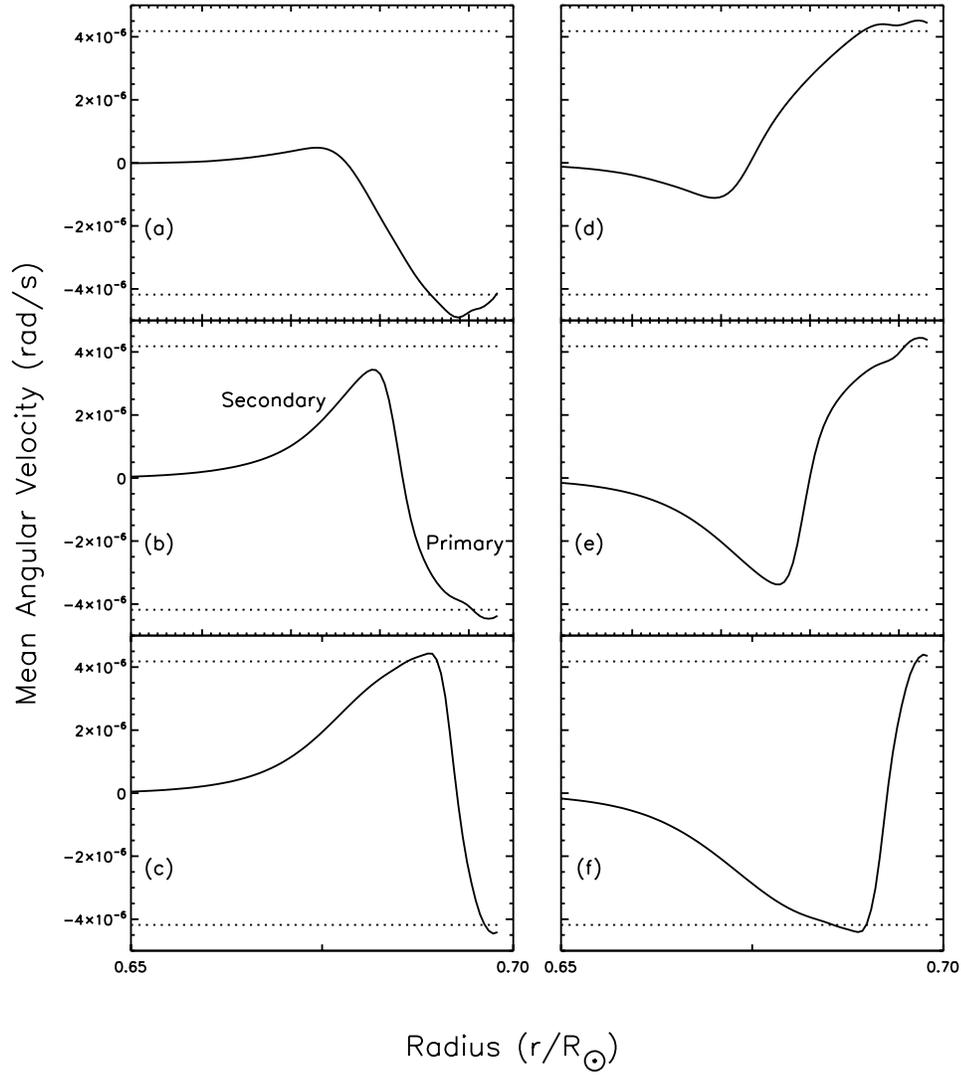}
\caption{Mean angular velocity at several times over one cycle for model M9.  Dashed lines represent the phase speed of the driven wave.}
\end{figure}
\clearpage
\begin{figure}
\includegraphics[width=6in]{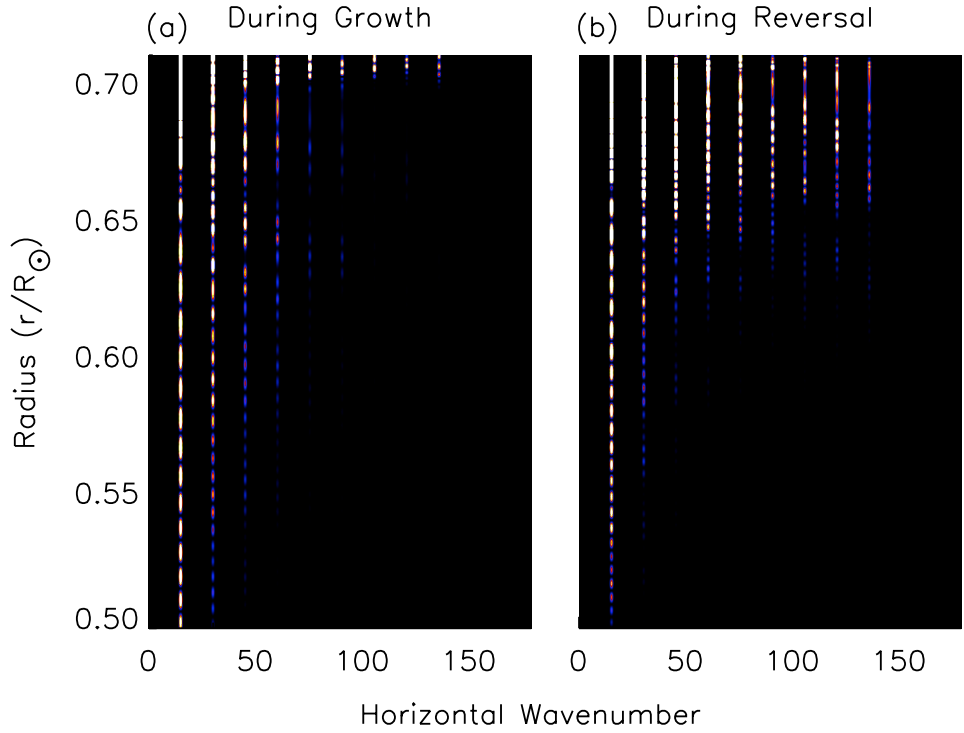}
\caption{Wave energy as a function of wavenumber and radius for model M9.  During growth of the secondary flow (left) wave energy is found primarily in the driven wavenumber (m=15).  During reversal of the primary flow (right) wave energy is found distributed amongst all possible modes.  The transfer of energy to higher harmonics allows rapid wave dissipation and associated angular momentum transfer.}
\end{figure}
\clearpage
\begin{figure}
\includegraphics[width=6in]{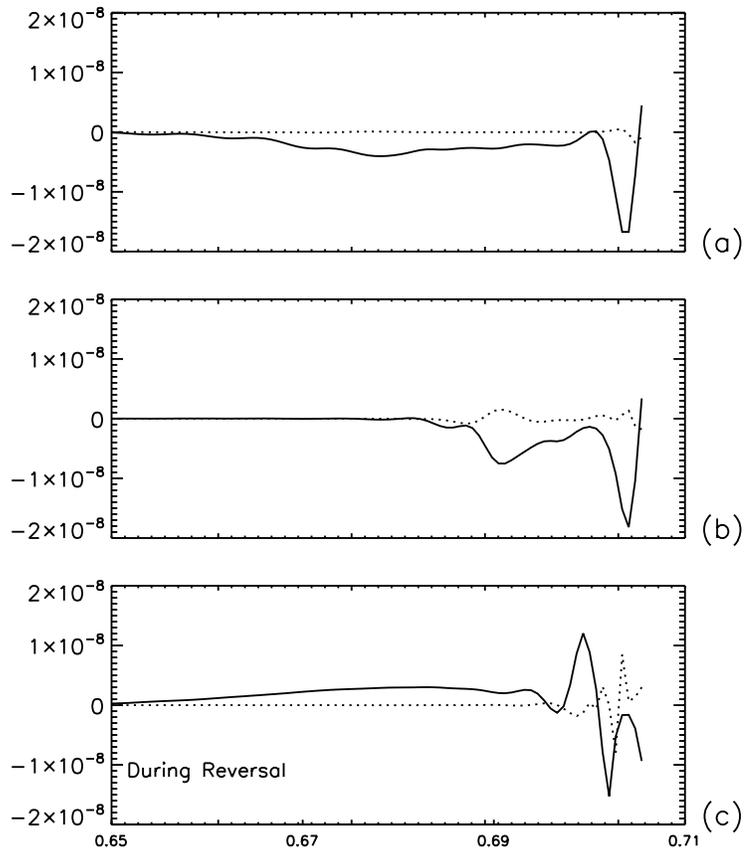}
\caption{One half cycle of model M11.  The solid line is the divergence of the nonlinear Reynolds stress (the second term LHS of equation 12, and the dotted line is the divergence of the viscous stress acting on the mean flow (the RHS of equation 12), both in $cm/s^{2}$.  The nonlinear term is ${\it always}$ larger than the viscous term acting on the mean flow.}
\end{figure}

\clearpage
\begin{figure}
\includegraphics[width=6in]{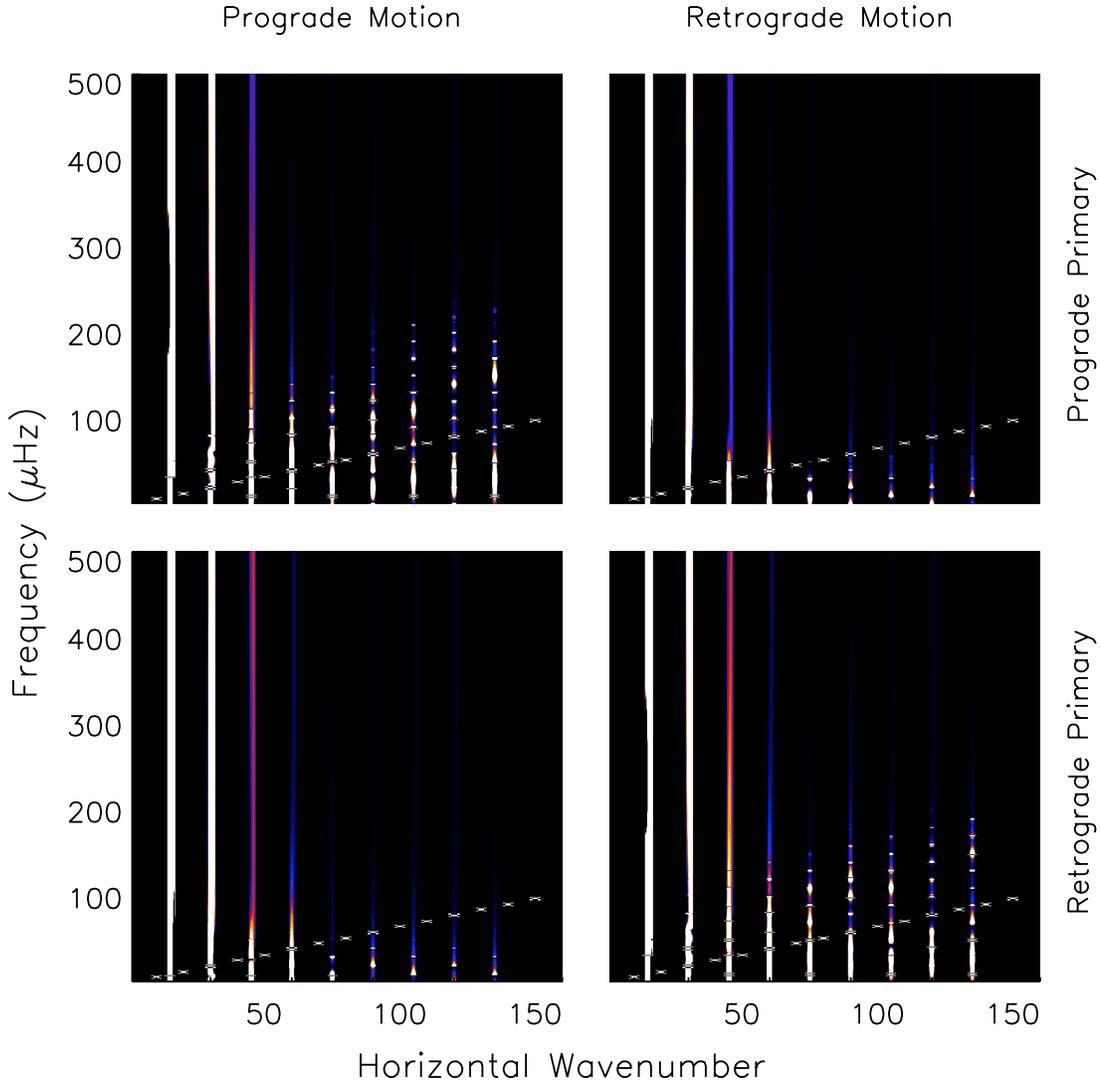}
\caption{Wavenumber (m)-Frequency (f) diagram of wave energy for prograde and retrograde waves at 0.69R$_{\odot}$ for model M2.  Top panels show this when there is a prograde primary mean flow at this radius; whereas the bottom panels are when a retrograde primary mean flow exists.  The diagonal line represents the mean angular velocity.  Wave energy below this line should be trapped by a critical layer.  This shows that waves of the same sense as the mean flow are doppler shifted to higher frequencies, so that significant wave energy is able to propagate past the critical layer.}
\end{figure}
\clearpage
\begin{figure}
\includegraphics[width=6in]{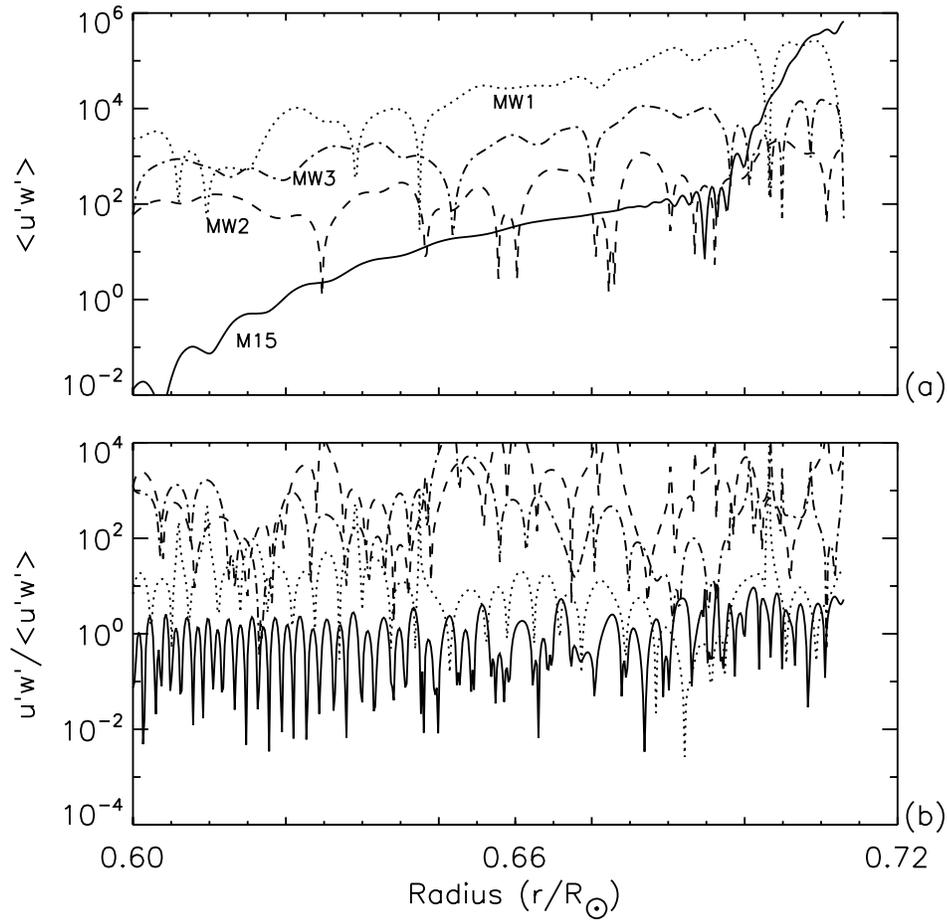}
\caption{(a) Horizontally averaged Reynolds stress (HARS) versus radius for models with multiple waves forced, compared with models with only a single (prograde and retrograde) wave forced (M15).  (b) Ratio of Reynolds stress at a randomly picked azimuth to the HARS.  The linetype is the same as in (a).  As multiple waves are forced, the ratio strays increasingly from one.}
\end{figure}
\clearpage
\begin{figure}
\includegraphics[width=6in]{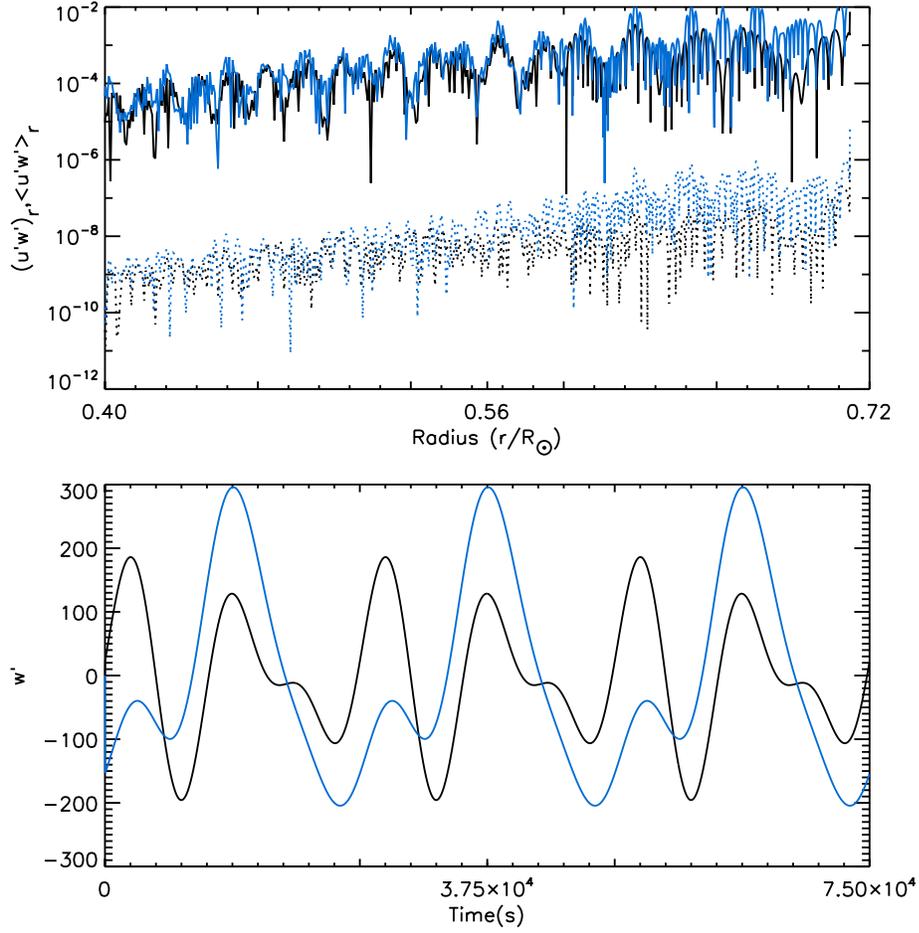}
\caption{(a) Radial derivative of the HARS (dotted lines) compared with the radial derivative of the non-averaged Reynolds stress (solid lines) with black representing model MC2 and blue representing model MC3, (b) radial velocity perturbation as a function of time at 0.69$R_{\odot}$ and a randomly chosen azimuth.  In both cases the inertial term is the same order as the nonlinear terms}
\end{figure}

\clearpage
\begin{figure}
\includegraphics[width=6in]{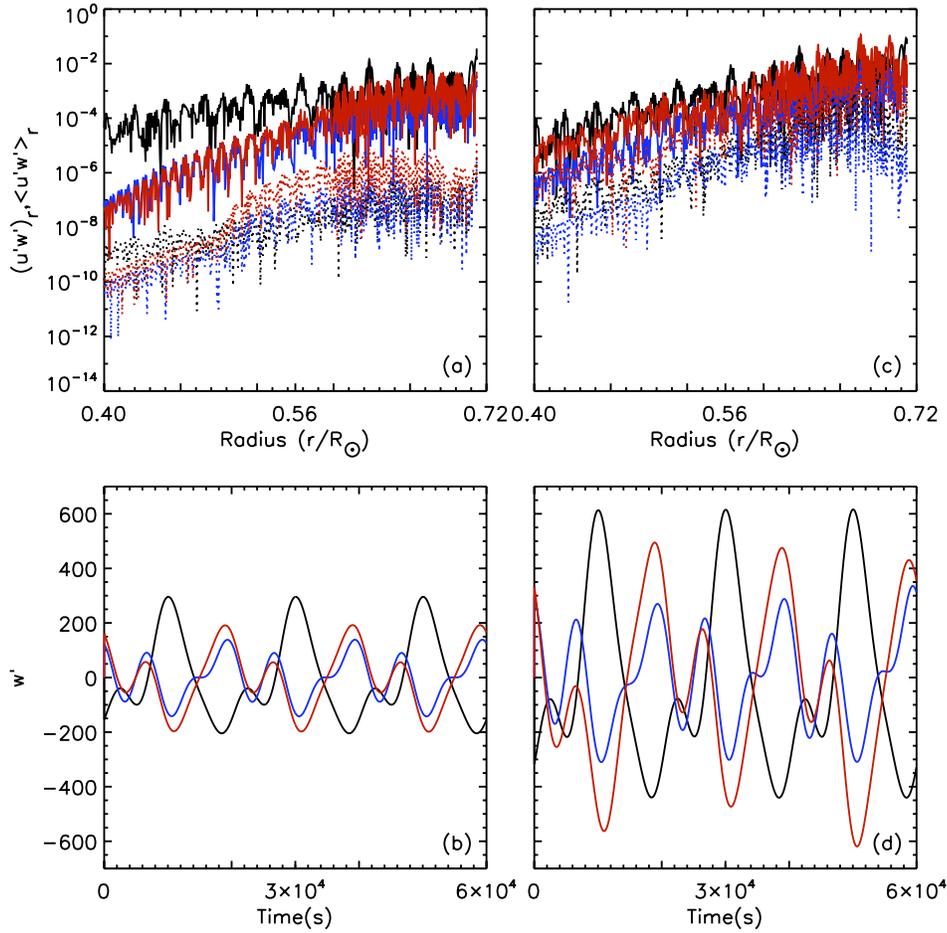}
\caption{Similar to figure 8, (a) shows radial derivative of the HARS (dotted lines) compared to the non-averaged Reynolds stress term (solid lines) for models MC3 (black), MSK5 (red) and MSF5 (blue); (b) radial velocity fluctuation as a function of time at 0.69 $R_{\odot}$ and a randomly chosen azimuth for the models portrayed in (a); (c) similar to (a) but for models MC4 (black), MSK6 (red) and MSF6 (blue); these models have a total energy input five times larger than those shown in (a); (d) same as (b) but for the models portrayed in (c).  In all cases the inertial term is the same order as the nonlinear terms.}
\end{figure}
\clearpage
\begin{figure}
\includegraphics[width=6in]{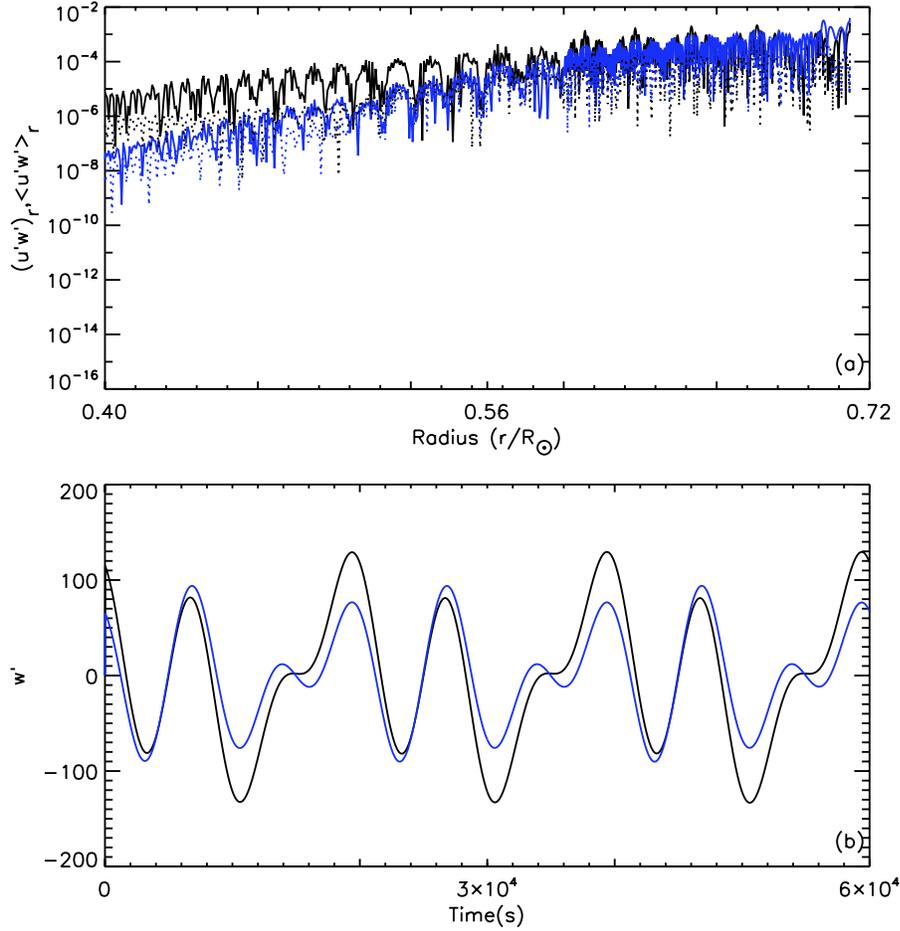}
\caption{Similar to Figures 8 and 9.  (a) Radial derivative of the HARS (dotted lines) compared to the non-averaged Reynolds stress term (solid lines) for models MF1 (black) and MF2 (blue).  (b) Radial velocity perturbations as a function of time at 0.69$R_{\odot}$ and a randomly chosen azimuth.  In both cases the non-averaged Reynolds stress is larger than the HARS and approximately equal to the inertial term.}
\end{figure}

\end{document}